**Title: Deep learning-based computed tomography (CT) derived body composition classifier for colorectal cancer patients.**

Eve Harling[1], Chattarin Pumtako[2], Bernd Porr[1], Donald C McMillan[2], Ross D Dolan[2]

1. James Watt School of Engineering, College of Science & Engineering, University of Glasgow, Glasgow, UK
2. Academic Unit of Surgery, School of Medicine, College of Medical Veterinary & Life Sciences, University of Glasgow, Glasgow, UK

Corresponding author:
Chattarin Pumtako
E-mail: C.Pumtako.1@research.gla.ac.uk
Room 2.51, New Lister Building,
Glasgow Royal Infirmary, Glasgow, United Kingdom, G31 2ER
ORCID: 0000-0002-7754-7229

## Abstract

**Background:** Accurate body composition analysis using Computed Tomography (CT) scans is essential for assessing skeletal muscle area (SMA) and skeletal muscle density (SMD), key markers of nutritional status in cancer patients. Conventional manual methods are labour-intensive and require specialist expertise, limiting their routine clinical use. Therefore, this study serves as a feasibility and pilot investigation to explore the potential of deep learning-based automated regression for body composition analysis within a clinical workflow.

**Methods:** Four deep learning architectures (AlexNet, UNet, GoogLeNet, and ResNet34) were trained to predict SMA, SMD, subcutaneous fat area (SFA), and visceral fat area (VFA) from CT scans of colorectal cancer patients. Systematic hyperparameter optimization identified the most accurate models, which were subsequently implemented in a web application for clinical use.

**Results:** GoogLeNet achieved the best performance, with a mean percentage error (PE) of 4.96% for SMA prediction, while AlexNet reached 8.12% for SMD. Independent testing demonstrated robust accuracy, correctly classifying body composition metrics in 80% of cases. The web application delivered rapid and consistent outputs, supporting integration into clinical workflows.

**Conclusion:** Optimized deep learning models, particularly GoogLeNet and AlexNet, can automate CT-derived body composition analysis with a Mean Percentage Error (PE) of 4.96% for SMA and 8.12% for SMD. These tools have the potential to streamline clinical practice by reducing the time and expertise required for manual segmentation. Further validation in larger, more diverse datasets is warranted.

## Introduction

Computed Tomography (CT) scans have emerged as invaluable tools in cancer management, providing detailed insights into patient anatomy that are crucial for prognosis and treatment planning. However, the full prognostic potential of CT-derived body composition analysis remains underutilized in clinical practice. This underutilization is largely due to the time-consuming nature of the analysis and the specialized expertise required to accurately interpret the data (1). Furthermore, the lack of access to advanced analytical tools, the high costs associated with these technologies, and insufficient training among healthcare professionals further limit the widespread adoption of these critical assessments (2). Historically, Body Mass Index (BMI) was used as a cancer prognostic; however, in light of rising obesity, this has now been shown to be an inferior measure when compared to CT-derived body composition analysis (3). BMI is an inadequate indicator because it fails to distinguish between lean muscle mass and adipose tissue. This limitation is particularly critical in colorectal cancer, where 'sarcopenic obesity'—a condition of severe muscle depletion masked by high body fat is strongly associated with poor quality of life and reduced survival (4, 5). In contrast, CT-derived metrics such as skeletal muscle area (SMA) and density (SMD) provide direct, precise quantification of a patient's true physiological status (2).

The importance of these metrics extends beyond simple body composition. For example, sarcopenia, which is characterized by the degenerative loss of muscle mass and function, is a common condition in cancer patients that is associated with poor clinical outcomes, including reduced survival rates (6, 7). Similarly, cachexia, which involves a disproportionate loss of skeletal muscle relative to fat, significantly impacts a patient's quality of life and overall prognosis (4). CT-derived metrics such as SMD can also indicate myosteatosis, a condition where muscle tissue is infiltrated by fat, further complicating a patient's clinical outlook.

One of the most significant CT-derived metrics is the Skeletal Muscle Index (SMI), which normalizes the SMA to a patient's height squared, providing a standardized measure of muscle mass that is strongly associated with survival in cancer patients. When combined with SFA, VFA, and SMD, these metrics can be used to calculate a CT-derived sarcopenia score (CT-SS), which serves as a powerful prognostic tool. The CT-SS is categorized into three levels, with higher scores indicating poorer prognosis. Despite their proven utility, the application of these measures in routine clinical practice remains limited due to the labour-intensive nature of manual analysis and the complexity of integrating these metrics into clinical workflows (8).

The advent of deep learning technologies offers a promising solution to these challenges. Deep learning, particularly convolutional neural networks (CNNs), has demonstrated remarkable success in automating complex image analysis tasks, making it feasible to incorporate detailed body composition assessments into routine clinical practice. By leveraging large datasets, these networks can learn to accurately predict key body composition metrics from CT scans, potentially transforming the way clinicians assess and manage cancer patients (5). Furthermore, the application of deep learning and optimization frameworks has been extensively explored across various domains of medical imaging and complex system modelling (9, 10). These studies provide a foundation for utilizing advanced computational architectures to enhance the accuracy of diagnostic tools, similar to the body composition classifier developed in this research. This study aims to evaluate the effectiveness of several deep learning architectures, including AlexNet (11), UNet (12), GoogLeNet (13), and ResNet34 (14) in predicting body composition metrics from CT scans of colorectal cancer patients. The ultimate goal is to develop a robust, clinically applicable tool that can enhance prognostic accuracy and improve patient outcomes in oncology.

## Methods

**Data Preparation:** This study utilized two primary datasets: one consisting of CT scans from the third lumbar vertebra (L3) of patients diagnosed with colorectal cancer at the Glasgow Royal Infirmary between 2008 and 2018, and another comprising anonymized patient information in a tabulated format. The CT scans were acquired using multi-detector scanners from major vendors (GE Medical Systems and Siemens Healthineers) at Glasgow Royal Infirmary. To ensure data harmonization across the wide acquisition period (2008–2018), all images were reconstructed using a standard soft tissue kernel with a slice thickness of 5 mm. Furthermore, only portal venous phase contrast-enhanced scans were selected to maintain consistency in Hounsfield Unit (HU) measurements across the dataset. Standardized windowing and intensity normalization were then applied during the pre-processing stage to minimize the impact of protocol heterogeneity on the deep learning models' performance. To ensure preprocessing consistency, all CT slices underwent Hounsfield Unit (HU) clipping within the range of -30 to +150 HU to optimize muscle tissue visibility. Following this, pixel values were normalized to ensure uniform input intensity across the dataset. These steps were strictly maintained to guarantee that the models received standardized input, regardless of the original CT scanner settings.

Initially, the CT scans were in DICOM format, which was incompatible with the TensorFlow framework used in this study. To address this, images were batch-converted to JPEG format using ImageJ (15). To mitigate the loss of Hounsfield Unit (HU) information, standardized windowing was applied to the DICOM data prior to conversion to normalize intensity ranges for soft tissue. This ensured that salient features, particularly those critical for skeletal muscle density (SMD) prediction, were preserved. For computational efficiency, images were subsequently resized from 512 x 512 to 256 x 256 pixels. A total of 574 scans were retained for analysis after a visual quality inspection for anisometry errors.

**Model Selection and Evaluation:** Four deep learning architectures—AlexNet, UNet, GoogLeNet, and ResNet34 were selected as benchmark models to evaluate their feasibility in medical regression tasks. While more recent architectures like Vision Transformers (ViTs) offer high performance, they typically require significantly larger datasets and higher computational power than was available for this study. The selected CNN-based models were chosen for their proven efficiency in extracting features from medium-sized medical datasets and their ability to operate effectively within the hardware constraints (NVIDIA GTX 1070) of this clinical feasibility study. These features make AlexNet (11) particularly well-suited for handling large-scale image data with relatively low computational costs. UNet (12) was included for its strength in biomedical image segmentation, especially its ability to accurately extract detailed features from images, which is critical for segmenting body composition metrics like SFA and VFA from CT scans. GoogLeNet (13) was selected for its innovative inception modules, which enable the model to capture multi-scale features within a single layer. This capability makes it particularly effective in distinguishing between different tissue types in complex CT images, while its use of global average pooling helps prevent overfitting. ResNet34 (14) was chosen for its depth and the use of residual connections, which help mitigate the vanishing gradient problem in deep networks, making it suitable for capturing intricate patterns in CT scan data, particularly for SMD and SMA predictions.

The models were evaluated based on their performance in predicting SFA, VFA, SMA, and SMD, using Mean Squared Error (MSE) and Percentage Error (PE) as metrics. Given that these output variables are continuous, regression models were constructed for each architecture. To perform numerical regression, the standard architectures were modified by replacing their original classification layers with a specialized regression head. This head integrated a global average pooling layer followed by a fully connected (dense) layer with a linear activation function, enabling the network to output continuous values for SMA, SMD, SFA, and VFA. Specifically, for UNet, the traditional expansive path typically used for pixel-wise segmentation was adapted by integrating a flattening layer and a regression head after the final convolutional block. This allowed the model to leverage its robust spatial feature extraction for direct numerical prediction rather than image mask generation.

The training process commenced with the setup of TensorFlow and Keras, the core libraries used for developing and fine-tuning the deep learning models. The CT scan data were carefully pre-processed and organized into a structured dataset, which included converting DICOM images to JPEG format and resizing them to 256×256 pixels to optimize for computational efficiency. To ensure an unbiased evaluation and prevent data leakage, data splitting was strictly performed at the patient level. This ensured that all scans belonging to a single individual were assigned exclusively to either the training, validation, or independent prediction sets. Following this strategy, an 80/20 split was applied for training and validation, while a separate group of 10 patients was held out as an independent prediction set to evaluate the final model performance. Hyperparameters, Initial hyperparameters, including a batch size of 16 (16), a learning rate of 0.0001, and 100 epochs, were established based on preliminary iterative testing to ensure stable convergence. Rather than an arbitrary selection, these values served as a baseline for subsequent systematic optimization using the Hyperband tuning algorithm. This approach allowed for an efficient exploration of the hyperparameter space—specifically filter numbers, dropout rates, and learning rates—to identify the configurations that minimized validation loss while respecting computational resource constraints.

**Data Augmentation and Training:** To improve the models' ability to generalize, data augmentation techniques including random horizontal flips, minor rotations (within ±10 degrees), and slight zooms were applied. These parameters were strictly controlled to ensure that the augmented images maintained anatomical plausibility and did not introduce unrealistic artifacts that could compromise the clinical integrity of the CT scans. This variability helped the models remain robust to minor differences in patient positioning and scanner alignment without distorting the underlying physiological metrics. The models were built using the Keras functional API, which provides flexibility in model architecture design. An 80/20 data split was used for training and validation, with the MSE recorded at each epoch as the loss function.

**Model Optimization:** To mitigate the risk of overfitting associated with a fixed 80/20 data split, this study employed Hyperband optimisation. Unlike a single-run training approach, Hyperband is a multi-fidelity resource allocation strategy that systematically evaluates hundreds of hyperparameter configurations. It utilizes a successive halving mechanism to discard unstable or underperforming models in early brackets, ensuring that only the most robust and generalizable architectures are selected for final evaluation. The stability of the final models is further evidenced by the consistent convergence observed in the learning curves (Figure 2).

Following the initial training, GoogLeNet demonstrated the highest accuracy in predicting SMA, making it the focus of further optimization using Hyperband tuning (17). Hyperband was selected for its efficiency in optimizing hyperparameters under limited computational resources. The key hyperparameters optimized for GoogLeNet included filter numbers, learning rate, and seed. Similarly, AlexNet, which performed well in predicting SMD, underwent optimization focusing on dropout rate, dense layer nodes, and learning rate. The specific search space and ranges of hyperparameters trialled during the Hyperband optimization process are detailed in Supplementary **Table S1**.

**Development of the CT-SS Classifier:** Post-optimization, the most accurate models—GoogLeNet for SMA and AlexNet for SMD—were used to develop a CT-SS classifier. The CT-SS was derived by classifying SMI and SMD as 'low' based on sex- and BMI-specific thresholds established in previous oncological studies. For SMI, the thresholds were set at $< 41\ cm^2/m^2$ for females and $< 53\ cm^2/m^2$ for males with a BMI ≥ 25, while a threshold of $< 43\ cm^2/m^2$ was used for males with a BMI < 25. For SMD, a threshold of < 33 HU for patients with a BMI ≥ 25 and < 41 HU for those with a BMI < 25 was applied. The CT-SS was then assigned as follows: a score of 2 (both low SMI and SMD), 1 (either low SMI or SMD), or 0 (both within normal ranges). This systematic categorization allows the predicted measures from the deep learning models to be converted into a clinically relevant prognostic tool.

**Web Application Development:** To enable the practical application of the developed models in clinical settings, a web-based application was created using Streamlit. This app allows clinicians to input patient data, including sex, height, BMI, and upload CT scans for analysis. The app predicts SMA and SMD using the trained models, calculates SMI, and provides a CT-SS score. The user-friendly interface ensures that the tool can be seamlessly integrated into clinical practice, facilitating quick and accurate assessments. The web application was developed using Python 3.10 and built upon the Streamlit framework (version 1.35) for the frontend interface. The core deep learning backend utilizes TensorFlow 2.15 and the Keras functional API. For reliable performance, external users are required to upload single-axial CT slices at the L3 vertebrae level in JPG, PNG, or JPEG format, with a resolution of 256x256 pixels. These specifications ensure that the reported performance metrics can be consistently replicated by external clinical users.

## Results

**Model Performance:** The performance of the four deep learning models was evaluated based on their ability to predict SFA, VFA, SMA, and SMD from CT scans using Mean Squared Error (MSE) and Percentage Error (PE) as the primary metrics. GoogLeNet demonstrated the strongest overall performance, achieving the lowest PE across multiple metrics; specifically, it predicted SMA with a PE of 4.96% (MSE: 450) and SMD with a PE of 8.04% (MSE: 34). AlexNet also showed strong results, particularly in predicting SMD, where it achieved a PE of 8.12% (MSE: 40). In contrast, UNet and ResNet34 exhibited higher error rates, with ResNet34 struggling most significantly with VFA prediction (PE: 12.02%, MSE: 5935).

**Figure 1.** Mean percentage error (PE) by model (1a) and by body composition metric (1b). GoogLeNet and AlexNet achieved the lowest overall PE compared with UNet and ResNet34.

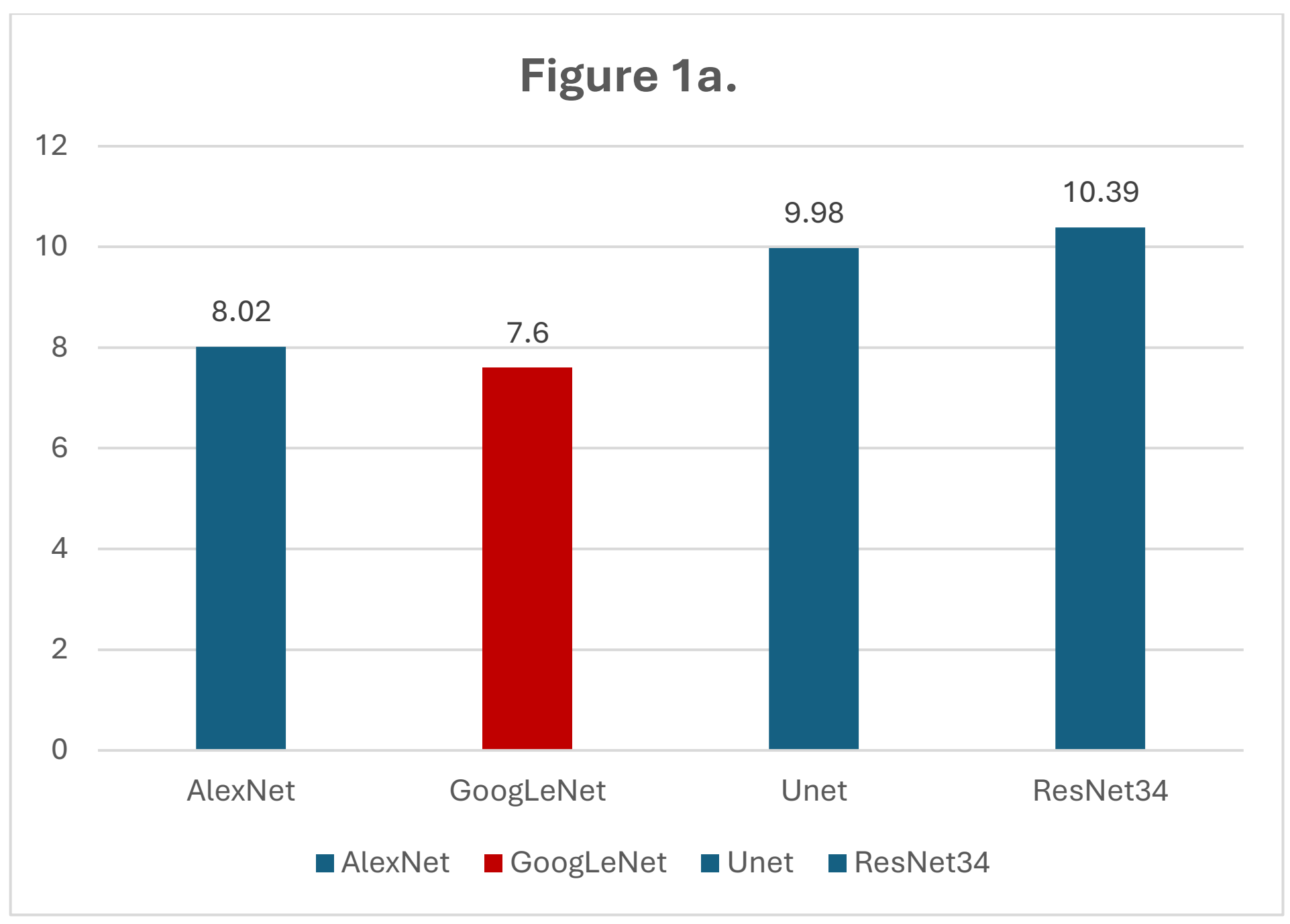

Figure 1a: Mean Percentage Error by Model

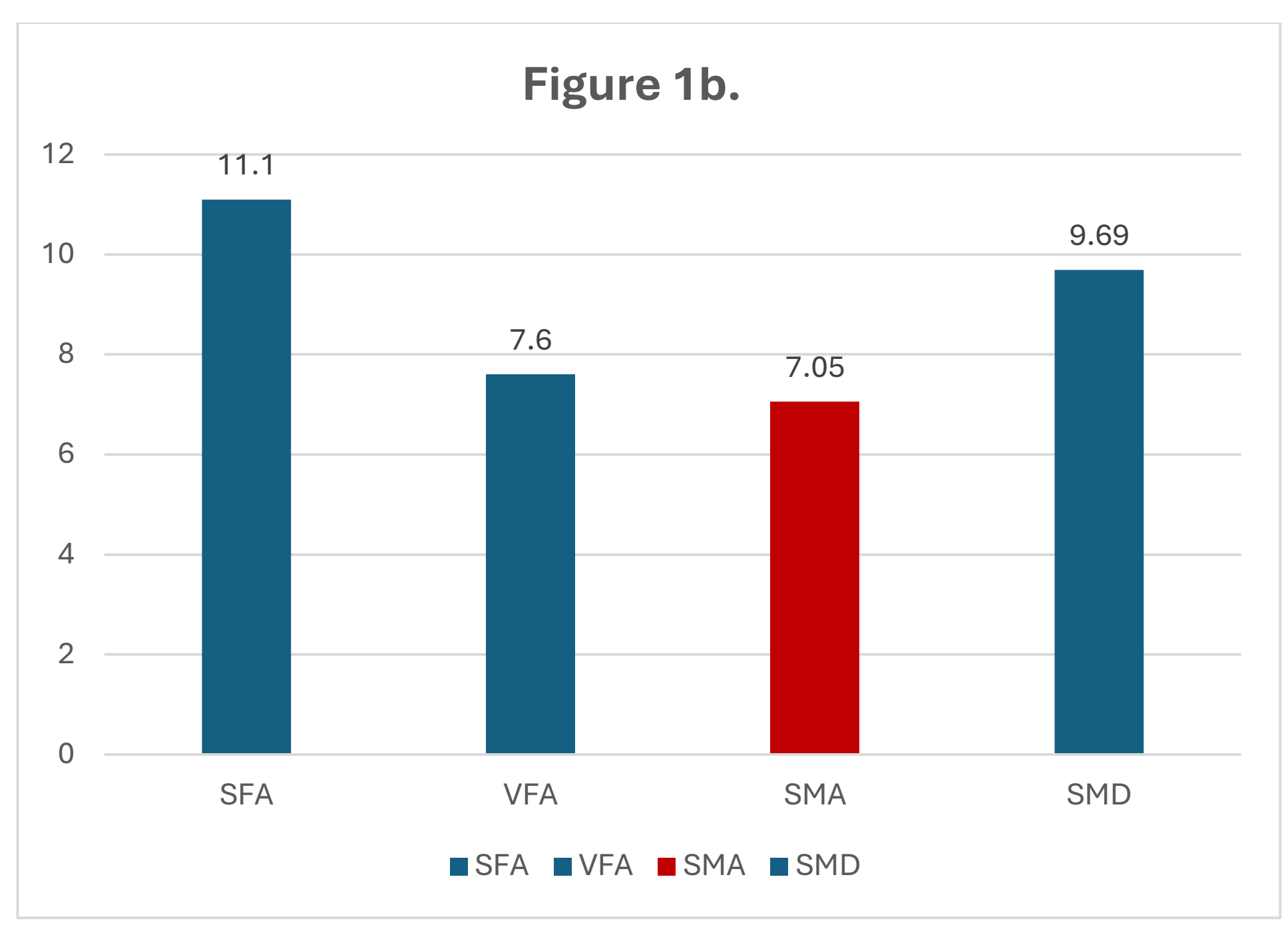

Figure 1b: Mean Percentage Error by Body Composition Analysis

Table 1. Comparison of the Different Models

| | AlexNet | | | GoogLeNet | | | UNet | | | ResNet34 | | |
|---|---|---|---|---|---|---|---|---|---|---|---|---|
| | MSE | AE | PE% | MSE | AE | PE% | MSE | AE | PE% | MSE | AE | PE% |
| SFA ($cm^2$) | 3412 | 58.41 | 8.16 | 6869 | 82.88 | 11.58 | 5935 | 77.04 | 10.76 | 9896 | 99.48 | 13.90 |
| VFA ($cm^2$) | 3998 | 63.23 | 9.52 | 1498 | 38.70 | 5.83 | 3493 | 59.10 | 8.90 | 3384 | 58.17 | 8.76 |
| SMA ($cm^2$) | 468 | 21.63 | 6.27 | 450 | 21.21 | 4.96 | 882 | 29.70 | 8.60 | 868 | 29.46 | 8.53 |
| SMD (HU) | 34 | 5.83 | 8.12 | 40 | 6.32 | 8.04 | 57 | 7.55 | 12.02 | 44 | 6.63 | 10.56 |

These results are summarized in **Table 1**. Comparative performance is further illustrated in **Figure 1a**, which shows GoogLeNet achieving the lowest overall PE at 7.60%, followed by AlexNet at 8.02%, while UNet and ResNet34 recorded PEs of 10.07% and 10.39%, respectively. Furthermore, **Figure 1b** highlights that SMA was the most accurately predicted metric across all architectures (mean PE: 7.05%), whereas SFA presented the greatest challenge (mean PE: 11.10%). To ensure experimental transparency and reproducibility, the final optimal hyperparameter configurations for the best-performing GoogLeNet (SMA) and AlexNet (SMD) models are provided in Supplementary **Table S2**."

**Time Efficiency:** The computational experiments were performed using an Intel(R) Xeon(R) CPU E5630 @ 2.53GHz and a GeForce GTX 1070 GPU. The combination of these hardware components allowed for efficient processing of the deep learning models. The time required per epoch for each model was recorded to assess computational efficiency. AlexNet completed each epoch in 1 second, making it the fastest model, GoogLeNet completed each epoch in 3 seconds, UNet completed each epoch in 12 seconds, whereas ResNet34 required 60 seconds per epoch, making it the slowest.

**Hyperparameter Optimization:** Hyperband hyperparameter tuning was applied to optimize the GoogLeNet model for SMA and the AlexNet model for SMD. After optimization, GoogLeNet's SMA prediction error was reduced from 6.15% to 4.96% after final epoch optimization, and AlexNet's SMD prediction error decreased from 9.28% to 8.12% following architectural refinement and epoch adjustment.

Table 2: Impact of Changes implemented

| Change Made | GoogLeNet SMA Model | | AlexNet SMD Model | |
|---|---|---|---|---|
| | PE% Achieved | Improvement % | PE% Achieved | Improvement % |
| Original | 6.15 | - | 9.28 | - |
| HyperBand Optimizing | 5.65 | -0.50 | 8.96 | -0.32 |
| Alternative End | 5.65 | 0 | 8.28 | -0.68 |
| Number of Epochs | 4.96 | -0.69 | 8.12 | -0.16 |
| Batch Size | 4.96 | 0 | 8.61 | 0 |
| Total | | 1.19 | | 1.16 |

Table 3: Comparison of categorisation of SMI and SMD as low or not low with scans split by factors affecting what threshold is used. Successful predictions have been highlighted in green.

| Group | Scan | SMI: Predicted (True) | SMD: Predicted (True) |
|---|---|---|---|
| Male, BMI<25 | Scan 1<br>Scan 4<br>Scan 6 | Not Low (Not Low)<br>Not Low (Not Low) | Low (Not Low)<br>Not Low (Low) |
| | | Low (Low) | Low (Low) |
| Male, BMI≥25 | Scan 3<br>Scan 5<br>Scan 7 | Low (Low)<br>Low (Low)<br>Low (Low) | Not Low (Not Low) Low (Low)<br>Not Low (Not Low) |
| | Scan 8 | Not Low (Not Low) | Low (Low) |

| | Scan 9 | Not Low (Not Low) | Low (Low) |
|---|---|---|---|
| Female, BMI<25 | Scan 2 | Not Low (Low) | Low (Low) |
| Female, BMI≥25 | Scan 10 | Not Low (Low) | Not Low (Not Low) |

**Stratified Performance Analysis: Table 3** summarizes the classification results stratified by sex and BMI. The analysis revealed that incorrect classifications for SMI were more frequent among female patients, while errors in SMD categorization were primarily observed in male patients with a BMI < 25 kg/m$^2$. These findings provide a granular view of the model's current performance across different demographic subgroups and highlight specific areas for further refinement.

**Figure 2.** Performance of GoogLeNet in predicting skeletal muscle area (SMA) across epochs (2a), and performance of AlexNet in predicting skeletal muscle density (SMD) across epochs (2b).

Figure 2a: GoogLeNet with SMA

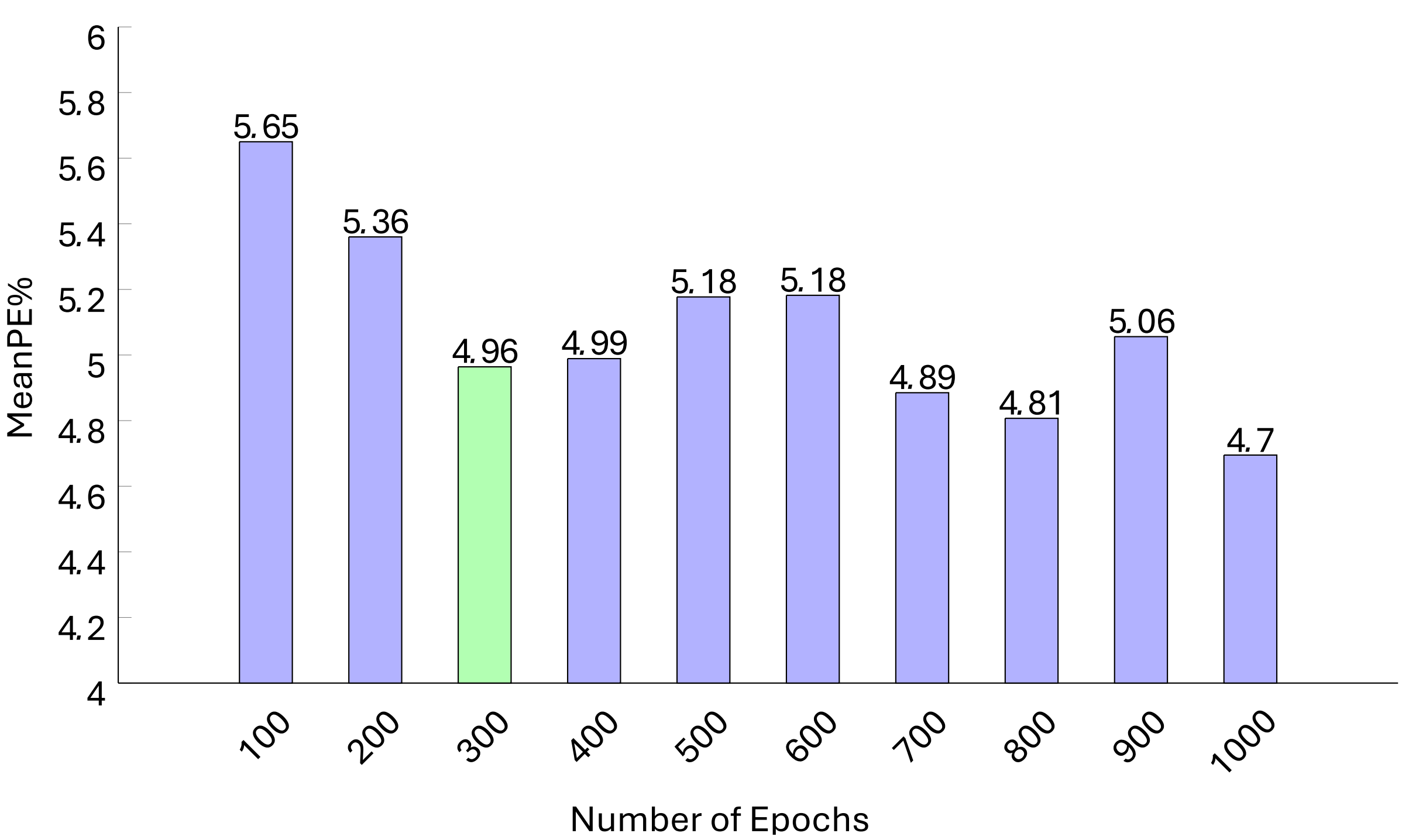

Figure 2b: AlexNet with SMD

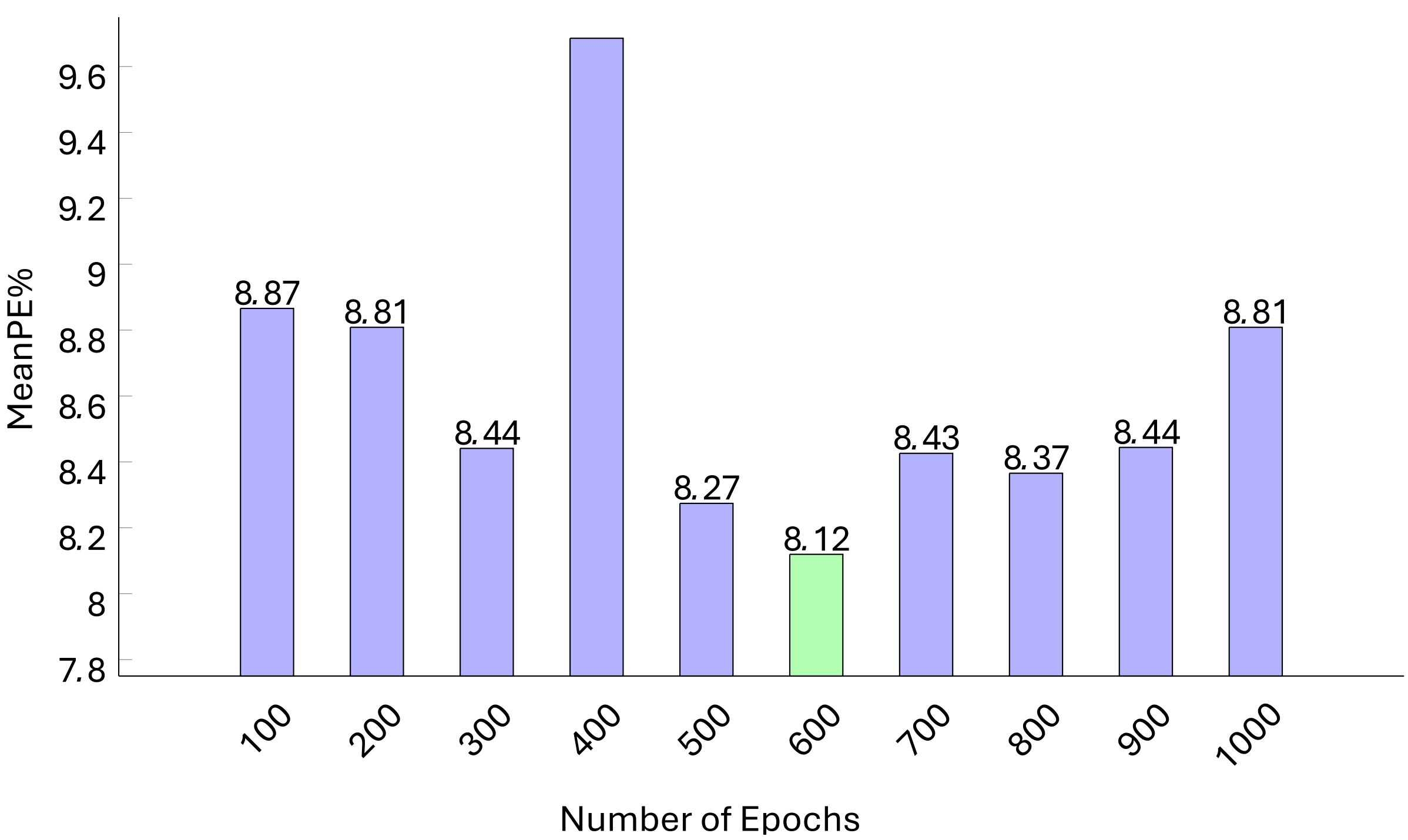

**Model Customization and Final Performance:** Further customization of the models was performed to enhance their performance. **Figure 2a** illustrates the change in PE when increasing the number of epochs for GoogLeNet with SMA. Although 1000 epochs exhibited the lowest PE, it demonstrated a high degree of unreliability and experienced interruptions during training due to out of memory (OOM) errors. A local minimum is found at 300 epochs that was consistent across multiple reruns and had no reliability issues. Given the selection of 300 epochs for GoogLeNet. **Figure 2b** shows the change in PE for AlexNet training with SMD at different epochs. The lowest PE value, 8.12%, achieved was at 600 epochs. There were no OOM errors encountered during training. Given the selection of 600 epochs for AlexNet, these values were determined through iterative testing and careful analysis of the models' learning curves. **Table 2** summarises the overall impact of these changes on the performance of the GoogLeNet and AlexNet models.

**Figure 3.** Scatter plots showing predicted versus true values for skeletal muscle index (SMI) (3a) and skeletal muscle density (SMD) (3b).

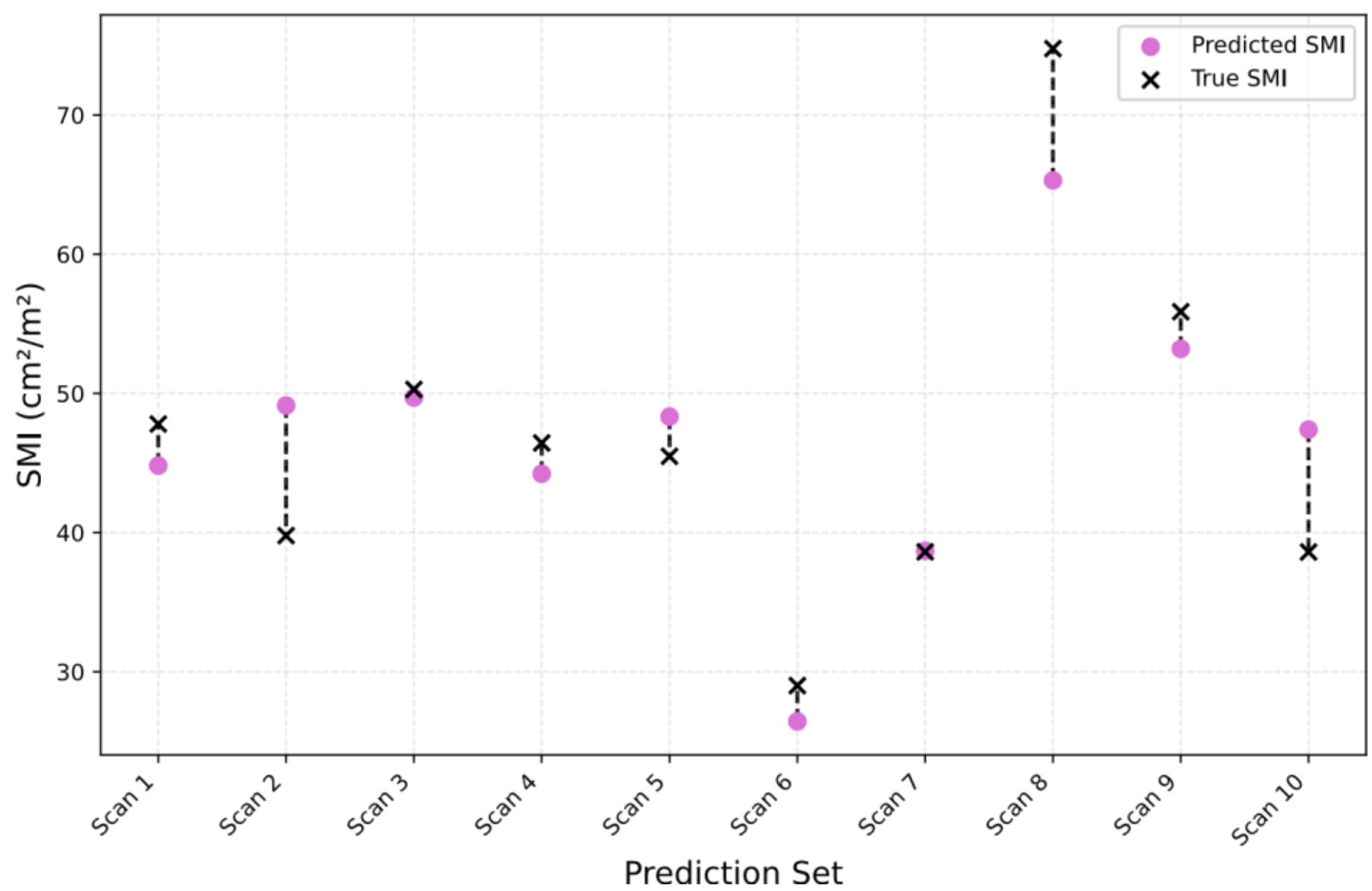

Figure 3a. Scatter Plot showing the Predicted and True SMI Values

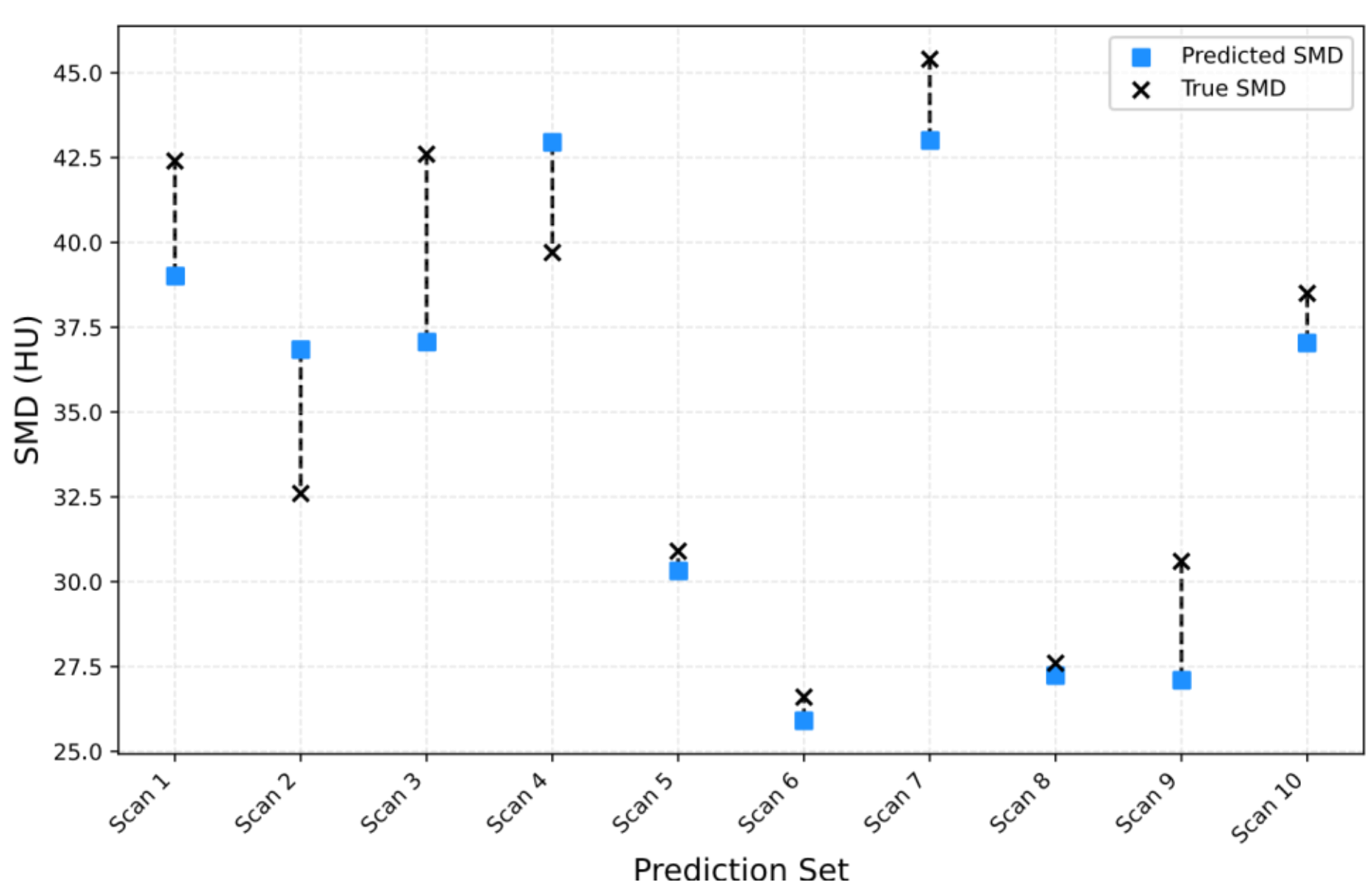

Figure 3b. Scatter Plot showing the Predicted and True SMD Values

**CT-SS Classifier Results:** The CT-SS classifier achieved an overall accuracy of 80%. To further evaluate its clinical utility, a class-wise analysis was performed. The classifier demonstrated high sensitivity for identifying high-risk patients (Score 2), while most classification errors (false positives) occurred in the intermediate Score 1 category. The balanced F1-score of 0.75 indicates a reasonable trade-off between precision and recall, though errors were more prevalent in specific subgroups, such as female patients for SMI and male patients with BMI < 25 for SMD. These findings suggest that while the 80% accuracy provides a strong foundation for automated screening, further refinement in demographic-specific thresholds is required to minimize false predictions in clinical practice.

**Figure 4.** Linear regression analysis between predicted and true values for (a) skeletal muscle index (SMI) and (b) skeletal muscle density (SMD).

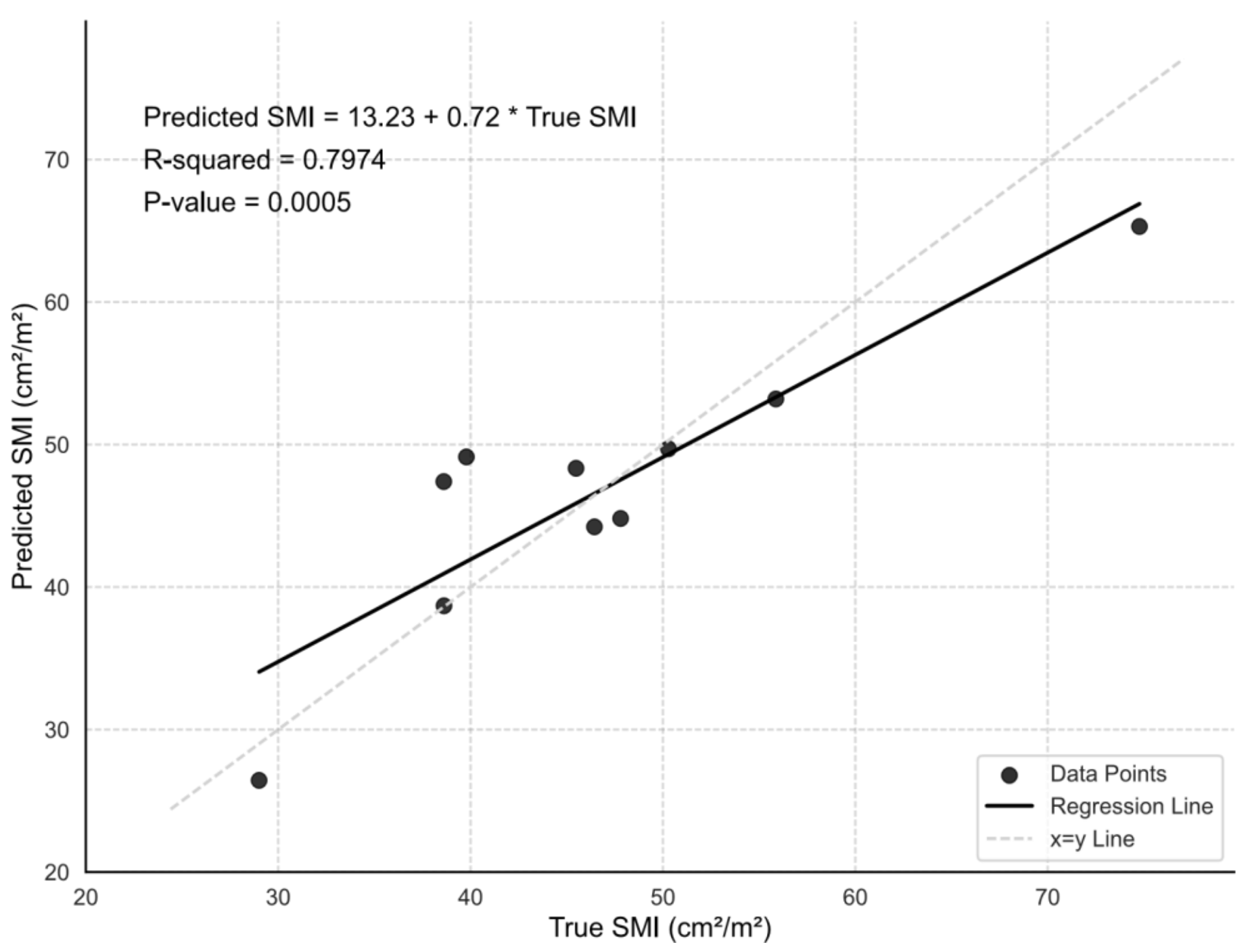

Figure 4a: Linear Regression Analysis between Predicted SMI and True SMI

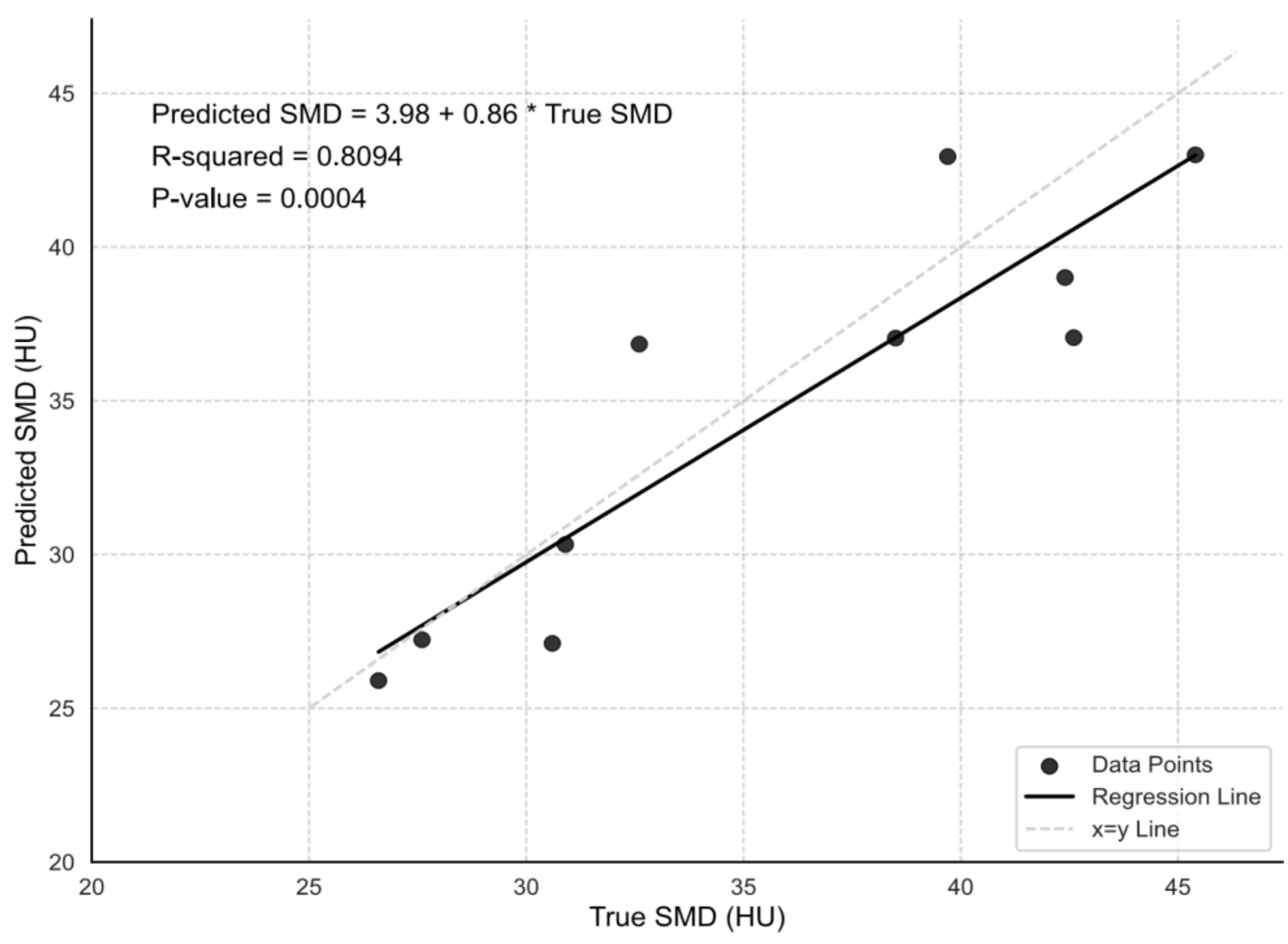

Figure 4b: Linear Regression Analysis between Predicted SMD and True S

The CT-SS classifier, developed using the optimized GoogLeNet and AlexNet models, was tested on a separate set of CT scans. The mean absolute error for SMI predictions was 2.299, with a relative mean error of 5.02% **Figure 3a.** The regression equation indicates that the associated p-value = 0.0005 for SMI predictions and true SMI **Figure 4a.** For SMD predictions the mean absolute error was 1.852, resulting in a relative mean error of 9.85% **Figure 3b**. The regression equation indicates that the associated p-value = 0.0004 for SMD predictions and true SMD **Figure 4b**. The model performance was characterized by strong clinical agreement and effect size rather than mere statistical significance. For SMI, the model achieved a high effect size ($R^2 = 0.7974$) with a Mean Absolute Error (MAE) of 2.299 $cm^2/m^2$. Similarly, SMD prediction demonstrated robust agreement ($R^2 = 0.8094$) with an MAE of 1.852 HU. While p-values were highly significant ($p < 0.001$), the focus remains on these agreement metrics, which confirm the model's reliability for clinical body composition assessment."

**Web Application Deployment:** A Model application was developed to integrate the trained models into a user-friendly interface for clinical use. The application allows users to input patient data, including sex, height, and BMI, and upload CT scans to receive predictions for SMA, SMD, and the CT-SS score.

**Discussion**

The deep learning models employed in this study were selected based on their balance between accuracy and computational efficiency, which are critical factors for practical clinical application. Among the architectures tested, GoogLeNet emerged as the most suitable for predicting SMA, while AlexNet was the top performer for SMD. These selections were driven by their performance metrics and computational demands during training. GoogLeNet's superior performance can be attributed to its innovative use of parallel layers with multiple filter sizes, which allows the model to process input data at various scales within a single layer. This design reduces the computational burden by incorporating global average pooling layers, which minimize the number of trainable parameters and help prevent overfitting. Additionally, the reliance on small, computationally inexpensive convolutions further enhances GoogLeNet's efficiency, making it comparable to AlexNet despite its deeper architecture.

AlexNet, although simpler and shallower than the other models, demonstrated the highest accuracy in predicting SMD. This is likely due to its compact architecture, which reduces the overall number of trainable parameters and accelerates the optimization process. The use of Max-Pooling and dropout layers further contributes to its computational efficiency by limiting the number of nodes and computations required per epoch. Interestingly, the relatively shallow depth of AlexNet may have allowed it to avoid overfitting on non-relevant features, thus maintaining high accuracy in SMD prediction.

Conversely, ResNet34 underperformed both in terms of accuracy and computational efficiency. The absence of layers designed to reduce computational costs, such as Max-Pooling or Global Average Pooling (GAP), combined with its substantial depth, likely contributed to its lower efficiency and accuracy. This underperformance suggests that ResNet34 may have encountered degradation issues, where the added depth did not translate into improved accuracy, potentially due to noise or irrelevant patterns in the data being learned by the model.

The prediction accuracy varied across the body composition metrics, with SMA and SMD being predicted more accurately than SFA and VFA. The PE for SMA was notably lower, especially with the GoogLeNet model, which achieved a PE of 4.96%. This performance underscores the model's robustness in capturing muscle-related features from CT scans. In contrast, SFA predictions were the least accurate, with a mean PE of 10.9%, proving particularly challenging for both GoogLeNet and ResNet34. The difficulty in distinguishing between SFA and VFA, given their similar appearances on CT scans, may have contributed to this higher error rate. Additionally, the edge detection required to separate SFA from the background might have been a limiting factor for these models. Despite these challenges, VFA was predicted with slightly better accuracy, indicating that the models could more reliably detect fat distributions closer to the visceral organs. SMD predictions, while more accurate than SFA, still presented challenges, particularly due to the nature of the metric, which relies on the brightness of the muscle in Hounsfield units. The percentage error for SMD averaged 10.44%, with AlexNet achieving the lowest error at 8.12%. This suggests that while the models are proficient at detecting muscle mass, further refinement is necessary to accurately capture muscle density.

The Hyperband optimization method proved effective in enhancing model performance, particularly for the GoogLeNet SMA model, where the percentage error was reduced by 0.459%. This improvement was achieved by fine-tuning the initial and final filter numbers, which increased the number of trainable parameters, leading to better accuracy without compromising computational efficiency. The learning rate, a critical parameter for balancing convergence speed and precision, was also optimized, contributing to the model's improved performance. For the AlexNet SMD model, the Hyperband tuning resulted in a smaller but significant improvement of 0.32%. The optimization focused on the dense and dropout layers, with the most substantial gain in accuracy (0.68%) achieved by replacing the fully connected, dense layers with those from GoogLeNet. This substitution not only enhanced accuracy but also reduced computational costs, demonstrating the effectiveness of combining elements from different architectures to leverage their strengths. The most considerable improvement for the GoogLeNet SMA model came from increasing the number of training epochs, which reduced the percentage error by 0.69%. However, the relationship between epoch count and error reduction was non-linear, with diminishing returns observed beyond certain thresholds, likely due to overfitting or noise in the dataset being misinterpreted as relevant features. Similar trends were observed for AlexNet, where the percentage error decreased up to 600 epochs but increased slightly thereafter, likely due to overfitting.

The CT-SS classifier, developed using the optimized GoogLeNet and AlexNet models, demonstrated an overall accuracy of 80% in classifying SMI and SMD as "low" or "not low." The classifier's performance, as indicated by the F1 score of 0.75, was reasonably balanced, though there is room for improvement, particularly in reducing false positives. The errors observed in classification were concentrated in specific patient subgroups, notably female patients for SMI and male patients with a BMI less than 25 for SMD. These findings suggest that further research is needed to explore sex- and BMI-related variations in model accuracy, potentially requiring tailored approaches for different demographic groups.

A key priority of this study was to ensure clinical practicality, production level reliability and accessibility. While state-of-the-art models like Vision Transformers (ViT) or nnUNet offer high performance, they often require high-end computational resources (specialized GPUs) and infrastructure (cloud-based servers) that are not typically available in routine clinical settings. By selecting established CNN architectures recommended for production such as GoogLeNet, U-Net and ResNet34 remain heavily utilized in the applied domain because they are thoroughly understood, robust, and demonstrate greater stability against out-of-training-distribution data (18). By selecting established CNN architectures recommended for production (see current TensorFlow tutorials), and adapting them for direct numerical regression, the present study developed a system that is both computationally efficient and capable of providing instantaneous results. This ensures that the tool can be integrated into standard clinical workflows and run on existing hospital hardware without the need for expensive infrastructure, which is a critical factor for widespread adoption in colorectal cancer care.

This focus on efficiency was directly realised in the development of the web application, which served as a proof of concept for integrating deep learning models into a user-friendly clinical tool. The app's performance was consistent, with predictions generated in under a second per scan, and the CT-SS predictions matched those calculated manually from the predicted values. From a practical deployment perspective within strict hospital IT environments heavy architectures typically require dedicated cloud-server installations, creating major bureaucratic hurdles and data-privacy risks. In contrast, the lightweight architectures utilized in the present study opens the door of running the entire algorithm right inside of a web browser, as achieved with tensorflow’s javascript implementation, which is highly desirable in terms of data protection as the input images and the results would stay on the local computer. However, the app's reliance on correct input data format and the lack of safeguards against erroneous inputs are notable limitations. Despite these issues, the app's intuitive interface and robust performance in handling valid data make it a promising tool for clinical use, provided further refinements are made to address its current limitations.

Beyond computational speed, the accuracy of the proposed regression-based approach remains highly competitive with complex segmentation frameworks. In comparing the present results with existing work, it was found that several studies have applied deep learning models to body composition analysis, particularly using abdominal CT scans. For instance, Hsu et al. (2021) developed a machine learning model to predict body composition metrics such as visceral fat area (VFA), subcutaneous fat area (SFA), and skeletal muscle area (SMA) from CT scans of patients with pancreatic cancer. However, their approach differs from the approach in this study in several key aspects: their study utilized a U-Net architecture to perform image segmentation (19), whereas the present work adopts a black-box approach, providing a numerical output for body composition metrics, including skeletal muscle density (SMD). Additionally, the current approach achieved a superior performance, particularly for SMA prediction, where GoogLeNet attained a percentage error (PE) of 4.96%.

Similarly, Dabiri et al. (2019) applied machine learning techniques to segment muscles and fat tissues from CT scans at the lumbar (L3) and thoracic (T4) levels in patients with colorectal cancer. Their approach also relied on image segmentation, focusing on muscle segmentation at L3, similar to the present investigation (20). However, the current study distinguishes itself by focusing on both segmentation and the prediction of numerical metrics, such as SMD and fat area, with AlexNet showing the highest accuracy for SMD prediction, achieving a PE of 8.12%. Although Dabiri et al. did not provide detailed performance metrics comparable to those reported here, it is clear that the proposed black-box numerical prediction approach offers a novel contribution, particularly for SMD prediction, which remains an underexplored area.

while previous work has primarily focused on image segmentation for body composition analysis, the present study presents a novel approach by providing numerical outputs for metrics such as SMA, SFA, VFA, and SMD. The current results indicate that the deep learning models tested, particularly GoogLeNet and AlexNet, offer competitive and in some cases superior accuracy to existing segmentation-based approaches, particularly in the prediction of SMA and SMD. This highlights the practical applicability of the proposed approach approach in clinical settings, where rapid and accurate predictions of body composition metrics are essential for assessing patient health, especially in individuals with conditions like colorectal cancer.

A primary limitation of this study is the reliance on a single-centre dataset from the Glasgow Royal Infirmary, which may restrict the generalizability of the models across diverse geographic and ethnic populations. As this study serves as a feasibility and pilot investigation to explore the potential of deep learning-based automated regression for body composition analysis within a clinical workflow, the primary focus was on establishing technical viability and real-time performance rather than definitive large-scale clinical validation. Consequently, while the independent prediction set (n=10) served as an initial functional test for the developed web application, its small size limits the statistical power of the evaluation. This was a necessary trade-off to ensure high-quality manual segmentations for each case during this pilot phase. Future work will prioritize multi-centre data acquisition and the implementation of k-fold cross-validation to provide more robust, unbiased performance metrics and ensure model stability across broader clinical settings. Additionally, while GoogLeNet's architecture was observed to be effective for SMA prediction, a formal ablation analysis was not performed within the scope of this feasibility study; such experiments will be included in future research to provide a more granular understanding of model accuracy.

To address the demographic biases identified in the stratified analysis (Table 3), future iterations of the model will explore adjustments such as stratified fine-tuning and the implementation of group-specific decision thresholds. By tailoring parameters specifically for female patients and individuals with a lower BMI, these refinements aim to ensure equitable performance and enhance clinical reliability across all patient profiles.

A further limitation is that this pilot study primarily utilised $R^2$ and Percentage Error (PE) to evaluate model performance. While these metrics showed high correlation with manual measurements, formal clinical agreement metrics such as the Intraclass Correlation Coefficient (ICC) and Bland–Altman analysis were not performed. Given the direct regression nature of the proposed models models, standard segmentation metrics like the Dice coefficient were not applicable. Future research involving a larger, multi-centre independent validation set will incorporate these statistical tests to definitively establish equivalence with the manual gold standard.

However, the study also highlights several areas for future research. There remains room for further optimization, particularly in improving the prediction accuracy of SFA and VFA, where the models exhibited higher PEs. Future work could explore the use of more advanced data augmentation techniques or the incorporation of transfer learning to further enhance model performance. Additionally, expanding the dataset to include a more diverse range of patient demographics and health conditions could improve the models' generalization capabilities. Moreover, real-world clinical validation is essential to ensure their robustness and reliability in everyday practice. The development of a user-friendly web application, as outlined in the study, represents a significant step toward this goal by providing clinicians with a practical tool for quick and accurate body composition analysis.

In conclusion, the results of this study demonstrate the feasibility of using deep learning models, specifically GoogLeNet and AlexNet, for accurate body composition analysis in clinical settings. The optimization processes applied have significantly enhanced the models' performance, making them more suitable for practical application. The study underscores the potential of these deep learning models to automate the analysis of CT-derived body composition metrics, which are crucial for assessing patient health, particularly in individuals with colorectal cancer. By accurately predicting these metrics, the models can assist clinicians in making informed decisions regarding patient prognosis, treatment planning, and monitoring the progression of conditions such as sarcopenia and cachexia. The models' ability to efficiently handle large-scale image data positions them as viable candidates for integration into clinical workflows.

**Acknowledgements:** The authors have nothing to report.

**Authors' contributions**

**EH** designed the work, acquired data, interpreting the results, revised the manuscript, Approved the final version. **CP** interpreting the results, drafted or revised the manuscript, Approved the final version. **BP** designed the work, revised the manuscript, Approved the final version. **DCM** designed the work, acquired data, interpreting the results, revised the manuscript, Approved the final version. **RDD** designed the work, acquired data, interpreting the results, revised the manuscript, Approved the final version.

**Ethics Statement**

This study was reviewed and approved by the West of Scotland Research Ethics Committee, Glasgow. All research was conducted in accordance with the Declaration of Helsinki. Written informed consent for participation was waived in accordance with national legislation and the requirements of the West of Scotland Research Ethics Committee because the study involved retrospective analysis of anonymized data collected during routine clinical care.

**Consent for publication**: Not applicable.

**Data availability:** The datasets generated and/or analysed during the current study are available from the corresponding author on reasonable request. The source code used for the analysis is publicly available at Zenodo (https://zenodo.org/records/10439107) (21)

**Competing Interests:** The author(s) declare no conflict of interest.

**Funding:** The author(s) received no specific funding for this work.

Table S1: Hyperparameter Search Space

| Hyperparameter | Values Trialled |
| --- | --- |
| **Random Seed** | 8, 16, 32, 42, 64 |
| **Initial Filter Number** | 32, 64, 128 |
| **Learning Rate** | $10^{-2}$, $10^{-3}$, $10^{-4}$, $10^{-5}$, $10^{-6}$ |
| **Dense Nodes (AlexNet)** | 100, 1000, 10000 |
| **Dropout Fraction (AlexNet)** | 0.4, 0.5, 0.6 |

Table S2: Final Optimal Hyperparameters

| Hyperparameter | Optimal GoogLeNet (SMA) | Optimal AlexNet (SMD) |
| --- | --- | --- |
| **Seed** | 32 | 8 |
| **Initial Filter Number** | 128 | N/A |
| **Final Filter Number** | 64 | N/A |
| **Learning Rate** | $1.0 \times 10^{-4}$ | $1.0 \times 10^{-4}$ |
| **Dropout Fraction** | N/A | 0.2 |
| **Dense Nodes** | N/A | 10 |
| **No. Epochs** | 300 | 600 |